# Long-term variations in solar meridional circulation from geomagnetic data: implications for solar dynamo theory

K.Georgieva[1][*] and B.Kirov[1],

[1] *Solar-Terrestrial Influences Laboratory at the Bulgarian Academy of Sciences, Sofia, Bulgaria*

**Abstract**

Geomagnetic activity has two main peaks in the 11-year sunspot cycle caused by two types of solar agents: coronal mass ejections and high speed solar wind streams, whose variations in number and intensity are related to the variations in solar toroidal and poloidal fields, respectively. From the long-term variations in the lag between the two aa-index peaks we derive the long-term variations in solar meridional circulation. We use these relations to test the flux-transport solar dynamo theory and to explain some well known characteristics of the solar cycle. Based on these relations, we give a prediction for the amplitude (125) and epoch of the maximum (April 2011) of sunspot cycle 24.

[*] Correspondence to: K.Georgieva (kgeorg@bas.bg)

# 1 Introduction

It has long been known that solar activity is the source of geomagnetic disturbances (see e.g. Cliver, 1994 for a review), so geomagnetic activity is often viewed as a proxy for solar activity. For quantifying the long-term variations in geomagnetic activity, aa-index is commonly used because this is the index with the longest available record - since 1868 (Mayaud, 1972), and for quantifying the long-term variations in solar activity the sunspot number is used available since 1611 (Hoyt and Schatten, 1998).

Geomagnetic activity as measured by aa-index has been steadily increasing since the beginning of the 20$^{th}$ century. Clilverd et al. (1998) studied all possible causes of this long-term increase and concluded that it could only be due to the variations in the interplanetary space and thus in the Sun. Lockhood et al. (1999), based on this increase in aa, speculated that the magnetic flux leaving the Sun has increased by a factor of 1.4 since 1964, and by a factor of 2.3 since 1901. Given the serious implications for the Sun's role in global climate, the above conclusion gave rise to severe controversy, until Cliver and Ling (2002) confirmed that this increase in the solar magnetic flux is real.

However, the correlation between aa-index and sunspot number in the 11-year solar cycle has been decreasing since the beginning of the 20$^{th}$ century, while the lag between the two has been increasing (Kishcha et al, 1999). As for the long-term correlation between aa-index and sunspot number, it was actually good only until a few decades ago, after which sunspot activity seems to be decreasing while geomagnetic activity continues increasing (Georgieva et al., 2005). The evolution of the correlation between solar activity and global terrestrial temperature is very much the same: until a few decades ago the global temperature closely followed the variations in sunspot numbers, and in the last decades sunspot activity seems to be decreasing while the temperature continues increasing. This gives arguments in favour of the anthropogenic influences on climate in the last decades (e.g. Solanki and Fligge, 1998); however, the similarity of the long-term changes in the correlations between sunspot number and terrestrial temperature, and between sunspot number and aa-index implies that they may have a common origin (Georgieva and Kirov, 2006).

Geomagnetic disturbances can be caused by agents originating from two different types of solar sources: open solar flux regions, such as coronal holes from which high speed solar wind causing recurrent geomagnetic activity originates, and closed solar field regions - sources of coronal mass ejections causing sporadic geomagnetic activity (Richardson et al., 2001). Corresponding to these two solar agents, there are two peaks in geomagnetic activity in the course of the 11-year solar cycle: one at sunspot maximum when the coronal mass ejections are most numerous and most intense, and another one on the declining to minimum phase of the sunspot cycle when polar coronal holes reach their maximum size.

Therefore, the changing correlation between sunspot number and aa-index can be due to either the changing lag between these two peaks, or to the changing of their relative importance, or to both.

Coronal mass ejections are related to the Sun's toroidal field, while coronal holes are related to the Sun's large-scale poloidal field (Gonzalez and Schatten, 1987; Bravo and Otaola, 1989). The cyclic transformation of the poloidal field into toroidal field and of this toroidal field into a new poloidal field with the opposite polarity - the solar dynamo, is the basis of solar activity. There are still many open questions regarding the mechanism of the solar dynamo, and very little is known about its long-term variations. It is the purpose of the present paper, based on the evolution of the correlation between sunspot and geomagnetic activity, to derive the long-term variations in the solar meridional circulation, a key element in the solar dynamo theory, and to compare the results with the existing theory and models. In a following paper we will use the evolution of the correlation between sunspot and geomagnetic activity to derive the long-term variations in the relative intensity of the solar toroidal and poloidal field, and to evaluate the implications for terrestrial climate.

In the next section we give a brief description of the flux transport dynamo theory (for a detailed review see for example http://solarphysics.livingreviews.org/Articles/). In § 3 we present our method of deriving the speed of the surface and deep meridional circulations, and study their long-term variations. In § 4 we investigate the role of the meridional circulation for modulating the amplitude and period of the sunspot cycle, and in § 5 we give a forecast for the next sunspot maximum. In the last section we summarize and discuss our results.

# 2 The solar dynamo

In sunspot minimum the Sun's magnetic field resembles the field of a bar magnet. In the regions around the two poles the magnetic field lines are open - stretching out into the interplanetary space. These regions are the polar coronal holes from which high speed solar wind originates and this high speed solar wind, when it hits the Earth's magnetosphere, gives rise to geomagnetic disturbances. As the coronal holes are long-lived structures, the Earth encounters these high speed streams each time when the coronal hole rotates into a position facing the Earth - that is, once every solar rotation. Therefore, geomagnetic disturbances caused by high speed solar wind are "recurrent" - reoccurring regularly every 27 days, the period of solar rotation.

As the solar cycle progresses, the solar differential rotation stretches the "poloidal" (meridional, or north-south) field in azimuthal (east-west) direction at the base of the solar convection zone (at about 0.7 R) giving rise to the toroidal component of the field - a process generally known as "Ω-effect" (Parker, 1955). The buoyant magnetic field tubes rise up, piercing the surface at two spots (sunspots) with opposite magnetic polarities. Due to the Coriolis force, the bipolar pair of spots is tilted with respect to the meridional plane with the leading (in the direction of solar rotation) spot at lower heliolatitude than the trailing spot. In each solar hemisphere, the leading spot has the polarity of the respective pole, and the trailing spot has the opposite polarity.

While this part of the dynamo - the transformation of the poloidal field into toroidal field - is believed to be clear, the physical mechanism responsible for the regeneration of the poloidal component of the solar magnetic field from the toroidal component (the so called α-effect) has not yet been identified with confidence (Charbonneau, 2005). Different classes of mechanisms have been developed. Recently the most promising seems to be the one based on the idea first proposed by Babkock (1961) and mathematically developed by Leighton (1969): late in the sunspot cycle, the leading spots diffuse across the equator where their flux is canceled by the opposite polarity flux of the leading spots in the other hemisphere (Dikpati and Charbonneau, 1999). The flux of the trailing spots and of the remaining sunspot pairs is carried toward the poles where it first cancels the flux of the previous solar cycle and then accumulates to form the poloidal field of the next cycle with polarity opposite to the one in the preceding cycle.

This mechanism requires a meridional circulation with a surface flow toward the poles where the poloidal flux accumulates, sinks to the base of the solar convective zone, and is carried by the counterflow there back to low latitudes to be transformed into toroidal flux and to emerge as the sunspots of the nest solar cycle. The near-surface poleward flux has been confirmed observationally from helioseismology (Hathaway, 1996, and the references therein; Zhao and Kosovichev, 2004; Gonzalez Hernandez et al., 2006), magnetic butterfly diagram (Ivanov et al., 2002; Svanda et al., 2007), latitudinal drift of sunspots (Javaraiah and Ulrich, 2006). The deep counterflow has not been yet observed, but has been estimated by the equatorward drift of the sunspot occurrence latitudes (Hathaway et al., 2003; 2004), though some studies question whether this is really an evidence of a material flow (Schüssler and Schmitt, 2004). There are some indications for long-term variations in both the surface and the deep meridional circulation (Hathaway et al., 1994; Javaraiah and Ulrich, 2006), but these variations are not known for sure. Such long-term variations in the meridional circulation would be very important for understanding the long-term variations in solar and geomagnetic activity, and we are deriving them in the following section.

## 3 Long-term variations in the meridional circulation

In this study we use the monthly sunspot numbers from ftp://ftp.ngdc.noaa.gov/STP/SOLAR_DATA/SUNSPOT_NUMBERS/, and the monthly aa-index (ftp.ngdc.noaa.gov/STP/GEOMAGNETIC_DATA/AASTAR/). The data series are smoothed by 12-month running mean, with a weight of 0.5 for the first and last points, and 1 for the rest of the points.

Fig.1 presents the smoothed monthly values of aa-index (solid line) and sunspot number (broken line) for the period 1868-2005. In the beginning of the period the peak in aa-index is very close to the sunspot maximum, and later on it begins exhibiting multiple maxima, with the principal one occurring on the declining branch of the sunspot cycle. We denote the time in months between the sunspot maximum and the following highest aa maximum on the declining phase of the sunspot cycle as "delay-1". The period between the sunspot maximum and the aa maximum on the declining phase of the preceding sunspot cycle is denoted as "delay+1"[*].

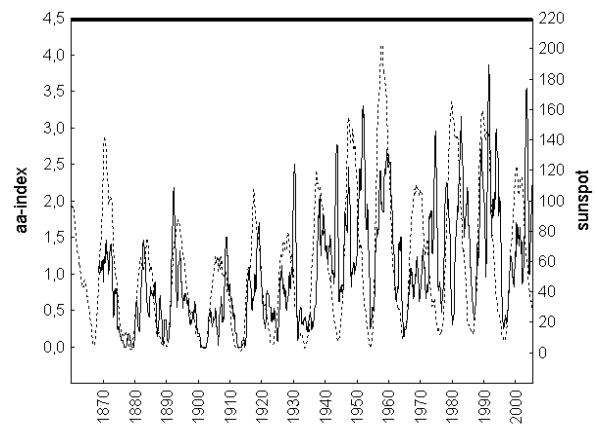

Figure 1. Smoothed monthly values of aa-index (solid line) and sunspot number (broken line) for the period 1868-2005.

### 3.1 Surface meridional circulation

Interpreting the Babcock-Leighton dynamo mechanism, the time between the sunspot maximum and the subsequent maximum in the solar poloidal field (delay-1) is the time it takes the surface meridional flow to carry the flux from low latitudes to the poles after sunspot maximum. In Fig.2a the values of delay-1 are plotted for consecutive sunspot cycles, 3-point running averages. Also shown for comparison is the sunspot number in the maxima of the respective cycles. From delay-1 we can estimate the speed of the surface meridional circulation, assuming that the distance traversed by the flow between sunspot maximum and the following aa maximum is from sunspot maximum latitude (~15°) to the pole (Fig.2b). The speed of the meridional circulation in the recent cycles, averaged over latitude from 15° to 90°, over time from sunspot maximum to the aa-index maximum on the sunspot cycle decline phase, and over the part of the convective zone involved in poleward circulation, was about 10 m/s. This is in good agreement with the results from helioseismology and magnetic butterfly diagrams cited above, showing latitude-dependent speed profile smoothly varying from 0 m/s at the equator to 20-25 m/s at midlatitudes to 0 m/s at the poles. (It should be noted that sunspot observations produce systematically lower values, see Javaraiah and Ulrich, 2006 for a discussion).

---

[*] (delay-1)$_n$ is the time between the sunspot maximum in cycle n and the highest aa-index maximum on the declining phase of cycle n; (delay+1)$_n$ is the time between the sunspot maximum in cycle n and the highest aa-index maximum on the declining phase of cycle n-1.

middle of the 20$^{th}$ century. As seen in Fig.2, the speed of the meridional circulation is obviously related to the amplitude of the sunspot cycle, leading it by one cycle. We will discuss this in Section 4.

### 3.2 Deep meridional circulation

Fig.3a shows the variations in delay+1, the time between sunspot maximum and the highest geomagnetic activity maximum on the decline phase of the previous sunspot cycle. This is the time for the high latitude poloidal flux formed after the previous polarity reversal to sink to the base of the convective zone, to be carried equatorward by the deep meridional circulation, to be transformed into toroidal field and to emerge in the active latitude belt as sunspots.

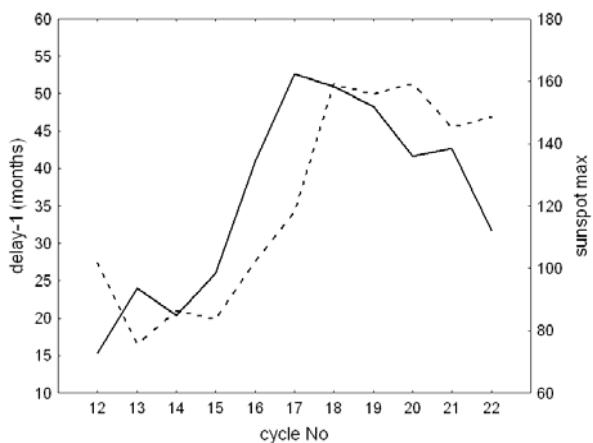

Figure 2a. Maximum sunspot number in consecutive sunspot cycles after 1868 (dashed line), and the time in months between the maximum sunspot number, 3-point running means, and the following highest aa-index maximum on the sunspot declining phase ("delay-1") (solid line), 3-point running means.

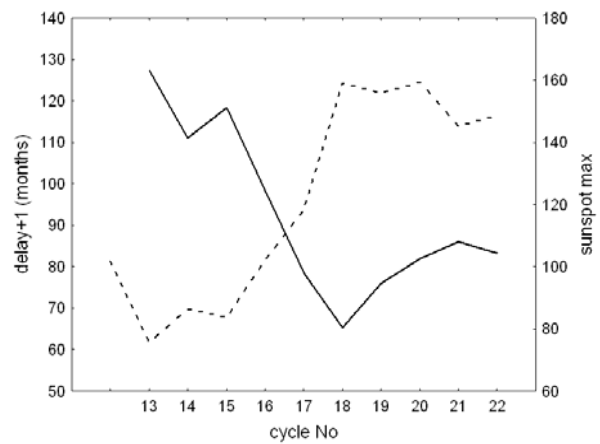

Figure 3a. Maximum sunspot number in consecutive sunspot cycles after 1868 (dashed line) and the time between th sunspot maximum, 3-point running means and the preceding aa-index maximum ("delay+1") - solid line, 3-point running means.

It is more difficult to estimate the speed of this deep circulation, because delay+1 includes not only the transport from the deep polar region to the deep active latitudes but also the vertical transport: sinking to the base of the solar convective zone at 0.7 $R_S$ at high latitudes, and emerging from 0.7 $R_S$ to the surface at low latitudes. (Some models assume it takes the meridional circulation two or three solar cycles to make a full turnover, e.g. Charbonneau and Dikpati, 2000, however in our opinion this means unrealistically low speed of the deep circulation; see the discussion below). While the time for buoyant emergence of flux tubes from the base of the convective zone to the surface is of the order of a month, the downward transport of the poloidal field at high latitudes depends on the relative importance of advection and diffusion, which is not known *a priori* (Yeates et al., 2007). We have calculated the speed of the deep meridional circulation in two limiting cases: assuming that the speed of the downward transport is equal to the speed of either the surface or the deep circulation, with the time for emergence of the flux at low latitudes set equal to one month in both cases. The difference between the two results is between 7 and 8

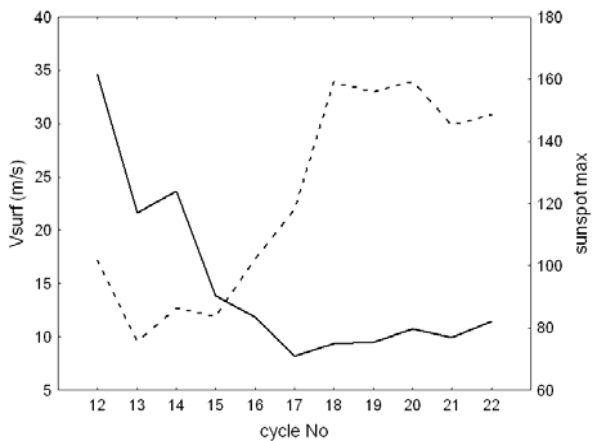

Figure 2b Maximum sunspot number in consecutive sunspot cycles after 1868 (dashed line), and the speed of the solar surface meridional circulation in m/s, 3-point running means.

An important feature evident in Fig.2b is that the speed of the surface meridional circulation has substantial long-term variations: from above 30 m/s in the end of the 19$^{th}$ century, to around 10 m/s since the

%. In what follows we use the values calculated with the speed of the downward transport equal to the speed of the deep meridional circulation (Fig.3b). In the recent cycles the speed of the deep meridional circulation which we derive from the above assumptions, average over latitude, over radial distance and over the sunspot cycle, is around 4 m/s. As shown by Hathaway et al. (2003), the speed varies with both latitude and time, decreasing from high to low latitudes and from the beginning toward the end of the sunspot cycle. Our result is in quite good agreement with their values between 1.5 and 2.7 m/s estimated from the sunspot butterfly diagram for sunspot maximum epoch and sunspot maximum latitudes. It is also in agreement with the model estimation of 3 m/s (Giles, 2000) consistent with mass conservation.

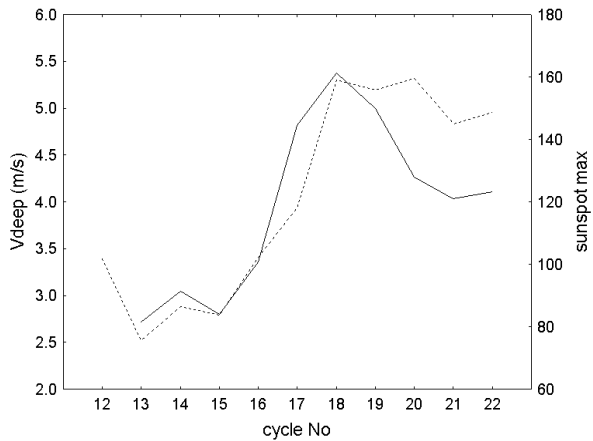

Figure 3b. Maximum sunspot number in consecutive sunspot cycles after 1868 (dashed line), and the speed of the solar deep meridional circulation, 3-point running means.

As in the case of the surface circulation, the speed of the deep meridional circulation also has long-term variations related to the secular solar cycle, from 2 m/s in the end of the 19$^{th}$ century to over 5 m/s in the end of the 20$^{th}$ century (Fig.3b), with higher speed corresponding to higher amplitude sunspot cycles. This will be discussed in Section 4.

### 3.3 Reversal depth

From Fig.2b and 3b it is evident that the speed of the surface meridional circulation is much higher than the speed of the deep return flow. This has to be expected from mass conservation because the density in the solar convective zone quickly increases with depth: 0.5% of the solar mass is contained in the upper half of the convective zone versus 2.5% in the lower half, so a slower deep flow is sufficient to balance the mass carried by the faster surface flow. Therefore, from the ratio of the speeds of the surface and deep circulations and from the mass distribution in the Sun we can estimate what part of the convective zone is involved in poleward and what in equatorward flow, if we know the radius to which the whole circulation system penetrates. Actually, there are no direct measurements of the penetration depth of the meridional circulation. Different models suggest values between 0.7 $R_S$ and 0.65 $R_S$, and even much deeper (Gilman and Miesch, 2004; Garaud and Brummell, 2007). Whether the meridional circulation is limited by the tachocline or penetrates some distance below it is important for the solar dynamo theory (Nandy et al., 2002), however we cannot estimate both the penetration depth and the reversal depth, so we assume that the penetration depth is known. The difference between penetration depth of 0.7 $R_S$ and 0.65 $R_S$ leads to a difference in the reversal depth of up to 0.03 $R_S$ with identical long-term variations in both cases. Fig.4 presents the long-term variations in the reversal depth of the solar meridional circulation for penetration depth of 0.7 $R_S$, with the mass distribution in the Sun taken from the Standard Solar Model BS2005-AGS,OP (Bahcall et al., 2005) available online at http://www.sns.ias.edu/~jnb/SNdata/solarmodels.html. Again, like the other studied parameters, the reversal depth exhibits well expressed long-term variations, from 0.84 $R_S$ in the low amplitude cycles at end of the 19$^{th}$ century to 0.74 $R_S$ in the strong cycles in the middle of the 20$^{th}$ century.

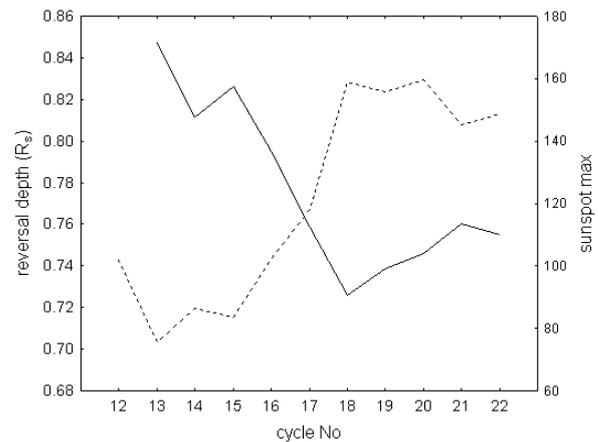

Figure 4. Maximum sunspot number in consecutive sunspot cycles after 1868 (dashed line) and depth where the circulation turns from poleward to equatorward (solid line), 3-point running averages.

## 4 Relation to the sunspot cycle, and implications for the solar dynamo theory

Surface circulation, deep circulation and reversal depth all exhibit long-term variations related to the long-term variations in solar activity (the secular, or Gleissberg solar cycle). Relations between the meridional circulation and the sunspot cycle amplitude and length are predicted by theory and are simulated by models, so the above results provide a test for the theory and the models.

### 4.1 Surface circulation and cycle amplitude

In the light of the flux transport dynamo, a slower surface circulation allows more leading-polarity flux to diffuse across the equator and to cancel with the opposite leading-polarity flux of the other hemisphere, therefore more trailing-polarity flux will be left to be carried to the poles where it will form a stronger poloidal field (Wang, 2004). Indeed, the correlation between the amplitude of the geomagnetic activity peak on the sunspot declining phase as a measure of the high latitude poloidal field ($aa_n$), and the speed of the surface meridional circulation between this aa-index peak and the preceding sunspot maximum ($Vsurf_n$) is negative, r=-0.841 with p=0.03 at lag=0 (Fig.5a).

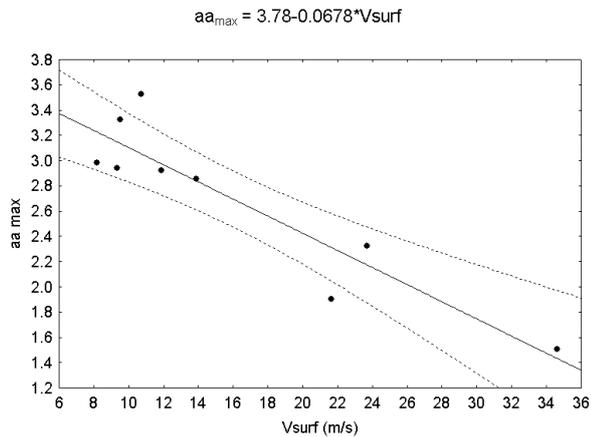

Figure 5a. Dependence of the maximum aa-index on the sunspot cycle decline phase on the speed of the surface meridional circulation.

Further, a stronger poloidal field will be transformed into a stronger toroidal field of the following sunspot cycle n+1. The correlation between the delay-1 and the amplitude of the following sunspot cycle is r =-.794 with p<0.02 (Fig.5b).

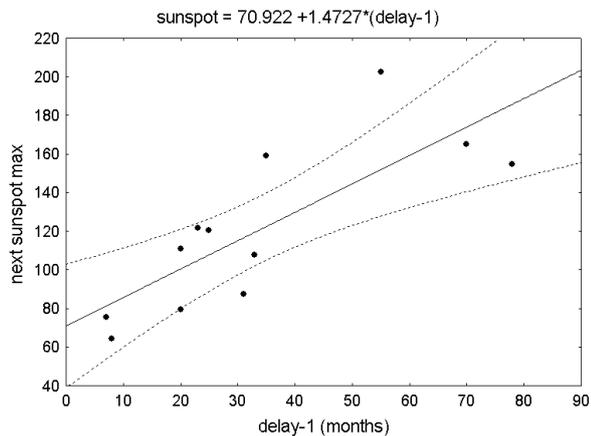

Figure 5b. Dependence of the maximum sunspot number in cycle n on the time between the previous sunspot maximum and aa-maximum on the decline phase of the previous solar cycle $(delay-1)_{n-1}$.

However, we find no correlation (r=-0.005) between the speed of the surface circulation and the amplitude of the *preceding* sunspot cycle as required by the model of Wang et al. (2002). Their simulations demonstrated that with constant surface circulation, the polar dipole stops reversing polarity. In order to maintain polarity reversal, a variable surface circulation is needed. By assigning values to Vsurf such that higher sunspot cycles are followed by faster circulation, they were able to stabilize the polarity oscillations around zero. The diamonds in Fig.5c indicate the values assigned to Vsurf in the simulation of Wang et al. (2002) while the filled circles are the values of Vsurf which we have estimated for the same cycles.

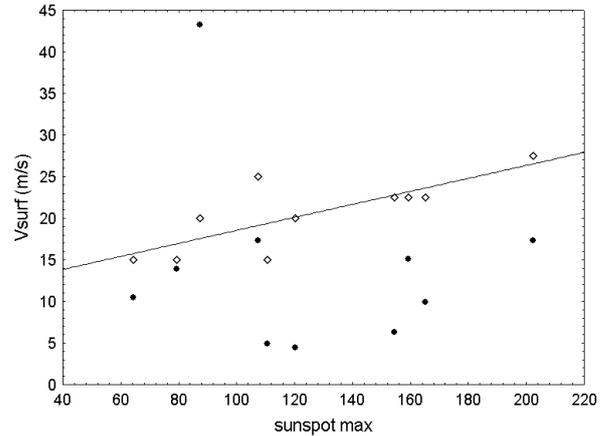

Figure 5c. Vsurf following sunspot maximum in the model of Wang et al. (2002) – diamonds, and estimated from delay-1 – filled circles.

Fig.5c demonstrates that a higher/lower sunspot maximum in one cycle is not followed by a faster/slower surface circulation in the same cycle. Instead, as seen from Fig.5b, a shorter/longer delay between sunspot maximum and the following aa-index maximum, respectively a faster/slower surface circulation in one cycle is followed by a lower/higher sunspot maximum in the next cycle. This relation has a predictive power: the amplitude of the following sunspot cycle can be predicted as soon as the maximum in geomagnetic activity on the sunspot declining phase is registered. In the § 5 we will use this relation to predict the amplitude of cycle 24.

## 4.2 Surface circulation and deep circulation

In models including variability of the meridional circulation (e.g. Yeates et al., 2007), it is tacitly assumed that faster/slower surface circulation means faster/slower deep circulation. Actually, as seen from Fig.2b and Fig.3b, while the speed of the surface meridional circulation is anti-correlated with the amplitude of the sunspot cycle, the speed of the deep circulation is positively correlated with it. Therefore the speeds of the surface circulation and of the deep circulation are anti-correlated, and faster/slower surface circulation means slower/faster deep circulation. For prognostic purposes it is more convenient to compare the delays rather than the speeds of the surface and deep circulation. The scatterplot of delay+1 versus delay-1 is shown in Fig. 6 with delay+1 lagging delay-1 by one cycle, and the correlation between (delay-1)$_n$ and (delay+1/)$_{n+1}$ is r=-0.69 with p=0.012.

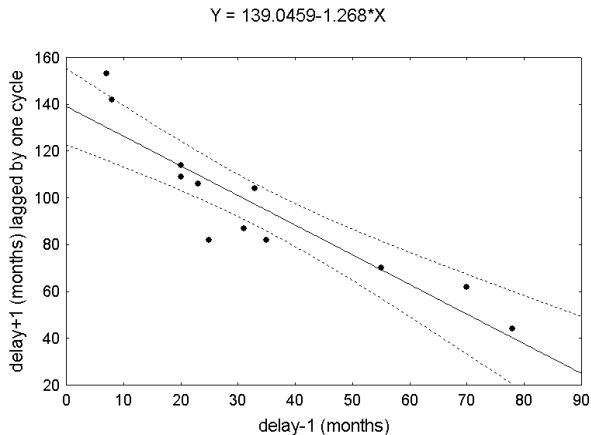

Figure 6. Dependence of the time from sunspot maximum in cycle n to aa-index maximum on the sunspot decline phase (delay-1)$_n$, and the time between this aa-index maximum and the following sunspot maximum (delay+1)$_{n+1}$.

There is no correlation (r=-0.1) between (delay-1)$_n$ and (delay+1)$_n$. This means that the speed of the deep meridional circulation is related to the speed of the surface circulation preceding it, but not to the speed of the surface circulation following it. This explains why there is no correlation between the amplitude of a sunspot cycle and the speed of the surface circulation following it as pointed out above: the amplitude of the sunspot cycle is related to the speed of the deep circulation preceding it, and there is no correlation between the speed of the deep circulation and the speed of the surface circulation following it.

### 4.3 Deep circulation and cycle amplitude

From Fig.3b it is evident that the speed of the deep meridional circulation preceding sunspot maximum is correlated to the magnitude of this maximum. The correlation is positive with r=0.72 and p<0.01 at lag=0 (Fig.7a) - a higher sunspot maximum appears after faster deep meridional circulation.

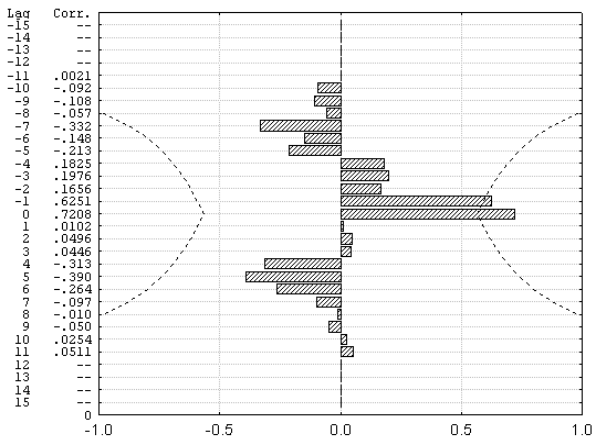

Figure 7a. Dependence of the maximum sunspot number in a cycle on the speed of the deep meridional circulation preceding it.

According to the model of Yeates et al. (2007), the dependence of the cycle amplitude on the circulation speed indicates whether the solar dynamo operates in "diffusion-dominated" or in "advection-dominated" regime: in diffusion-dominated regime (relatively high diffusivity and relatively low circulation speed), a higher circulation speed means less time for diffusive decay of the poloidal field during its transport through the convective zone, leading to more generation of toroidal field and hence a higher cycle amplitude. In advection-dominated regime (relatively high circulation speed and relatively low diffusivity), diffusive decay is less important and a higher circulation speed leads to a lower cycle amplitude because there is less time to induct toroidal field in the tachocline. Therefore, this result implies that in the last 140 years the solar dynamo has been operating in diffusion-dominated regime.

### 4.4 Deep circulation and cycle rise time

The dependence of the cycle amplitude on the speed of the preceding deep circulation in the diffusion-dominated mode explains why the rise time of a cycle anti-correlates with the amplitude of the cycle, the so-called Waldmeier rule (e.g. Hathaway et al., 1994; Hathaway and Wilson, 2004; Cameron and Schüssler, 2007; Schüssler, 2007): the rise time is fully included in delay+1, the time it takes the deep meridional circulation to carry the flux from the poles to sunspot maximum latitudes (r=0.85 with p<0.01, Fig.7b), which is inversely proportional to the speed of the deep meridional circulation, and in the diffusion-dominated regime the speed of the deep meridional circulation is proportional to the amplitude of the cycle.

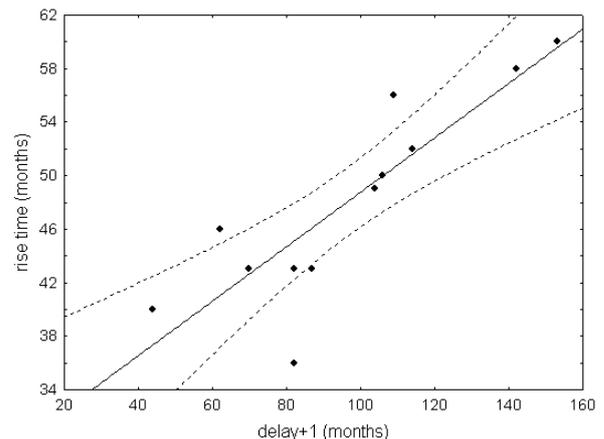

Figure 7b. Dependence on the rise time from minimum to maximum of cycle n on the time between sunspot maximum n and the previous aa-index maximum (delay+1)$_n$.

### 4.5 A case of advection-dominated regime: the Maunder minimum

There are no direct measurements of geomagnetic activity in the period of the Maunder minimum, however, we can instead derive the times of the maximum solar poloidal field from cosmogenic isotopes whose variations reflect the variations in the galactic cosmic rays modulation. The heliospheric modulation of galactic cosmic rays depends on the open solar magnetic flux through the solar source surface which is related to the global dipole component of the solar magnetic field (Usoskin et al., 2003). Here we use the years of minima in $^{10}$Be during the Maunder minimum (Beer et al., 1998) as the years of maximum poloidal field, the years of maxima in the group sunspot numbers (Hoyt and Schatten, 1998) as the years of maximum toroidal field. Fig.8a presents the estimated surface and deep circulation, and Fig.8b the relation between sunspot activity and the speed of the deep circulation.

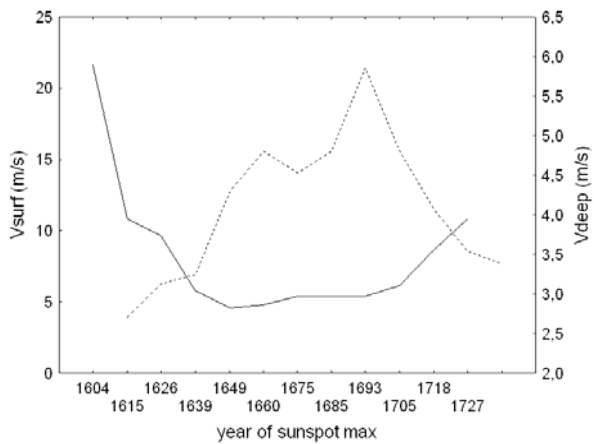

Figure 8a. Speeds of the surface (solid line) and deep (dashed line) meridional circulation during the Maunder minimum; 3-point running means.

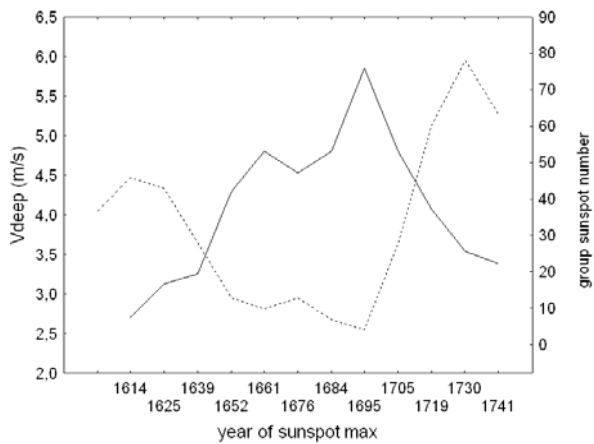

Figure 8b. Speed of the solar deep meridional circulation (solid line) and maximum group sunspot number in consecutive sunspot cycles (dashed line) during the Maunder minimum; 3-point running means.

In contrast to the recent cycles (Fig.3b), the correlation between the speed of the deep circulation and the amplitude of the sunspot cycle is negative: the faster the deep circulation, the lower the cycle amplitude (Fig.8c). This means that the solar dynamo operates in advection dominated regime which occurs when the deep circulation speed is relatively high and the diffusivity is relatively low, so increasing the speed means less time for the toroidal field to be generated in the tachocline (Yeates et al., 2007).

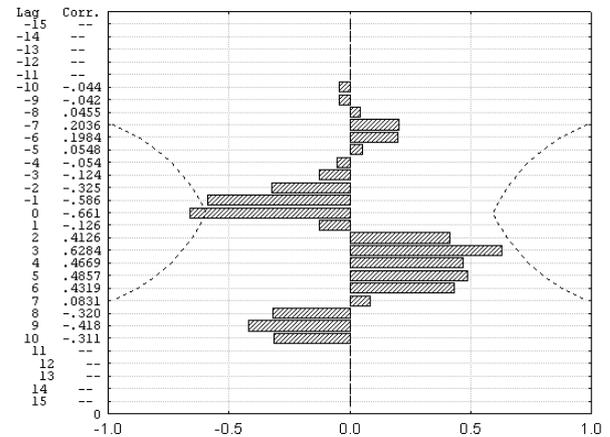

Figure 8c. Dependence of the maximum sunspot number on the speed of the deep meridional circulation during the Maunder minimum.

A comparison with Figs. 2b and 3b shows that both the 20$^{th}$ century secular solar maximum and the Maunder minimum correspond to periods of slow surface circulation and fast deep circulation. Further, it can be seen that while there is no real difference in the deep circulation speed in the two periods, the surface circulation is persistently slower during the Maunder minimum than during the 20$^{th}$ century secular solar maximum, and correspondingly, the reversal depth is lower (Fig. 8d, compare with Fig.4).

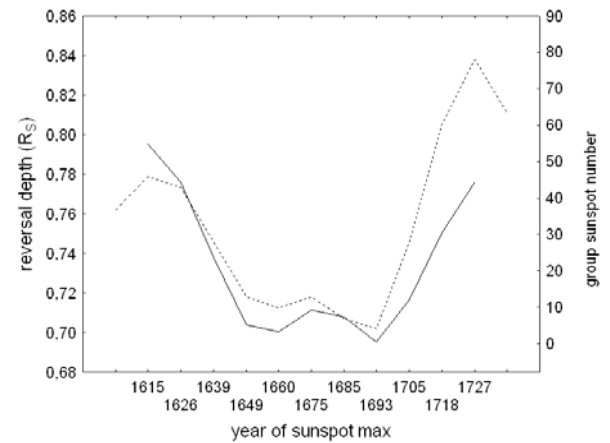

Figure 8d. Maximum group sunspot number in consecutive solar cycles during the Maunder minimum (dashed line) and reversal depth where the circulation turns from poleward to equatorward (solid line); 3-point running averages.

As Yeates et al. (2007) have shown, the change from positive to negative correlation between the speed of the deep circulation and the cycle amplitude can occur with increasing speed when the diffusivity is low. Though the depth profile of the diffusivity is not known, it is believed that in the lowermost part of the convective zone diffusivity quickly decreases with depth. Therefore lowering of the reversal depth means moving the deep equatorward circulation to deeper layers with lower diffusivity where the advective transport becomes more important than diffusion.

Using reconstructions of aa-index of geomagnetic activity since 1619 (Nagovitsyn, 2006), an estimation can be made of the variations in the solar meridional circulation in the last 4 centuries. It appears that the Maunder minimum in the 17$^{th}$ century and the following three secular solar maxima, in the 18$^{th}$, in the 19$^{th}$ and in the 20$^{th}$ centuries, all corresponded to minima in the speed of the surface meridional circulation, maxima is the speed of the deep meridional circulation, and consequently minima in the estimated reversal depth of the meridional circulation cell. However, in the period of the Maunder minimum, the surface poleward circulation was still slower, and the reversal depth was still deeper. We can therefore speculate that a grand minimum in solar activity possibly occurs when the solar dynamo is set up for a secular maximum (decreasing speed of the surface circulation, increasing speed of the deep circulation, lowering of the reversal depth), but for some reason the surface circulation slows down too much, and the deep return branch of the meridional circulation lands in layers with diffusivity below a certain threshold. It is not known what causes these changes, but they are probably related to the solar dynamics: it was established that during the Maunder minimum, the solar radius was bigger than in the contemporary epoch, and solar rotation was slower and more differential (Ribes et al., 1988).

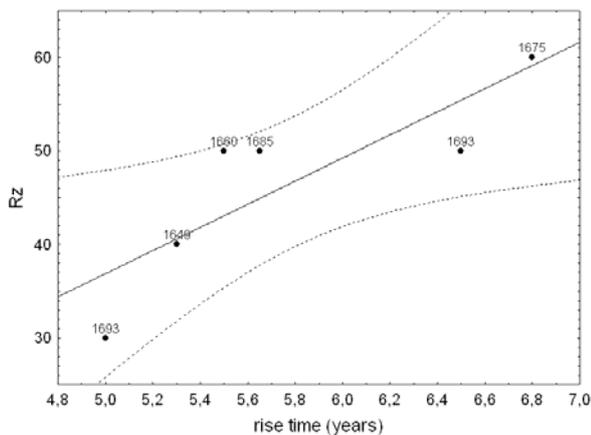

Figure 8d. Dependence of the maximum sunspot number in a cycle on the rise time from sunspot minimum to sunspot maximum during the Maunder minimum.

Finally, the Maunder minimum period can provide the test for the explanation of the Waldmeier rule relating the sunspot cycle amplitude to the cycle rise time, needed to prove the validity of the Babcock-Leighton type dynamo model (Schüssler, 2007). As predicted by the model of Yeates et al. (2007) and confirmed by the results shown in the previous subsections, in the diffusion-dominated regime the cycle amplitude is positively correlated to the speed of the deep meridional circulation which determines the rise time of a cycle: the higher the speed of the deep circulation, the shorter the rise time, and the higher the sunspot maximum. In the advection-dominated regime, the cycle amplitude is anti-correlated to the speed of the deep meridional circulation, so if this explanation is right, the rise time should be positively correlated to the amplitude of the cycle: the faster the deep circulation, the shorter the rise time, the lower the cycle. Fig.8d, based on data from the reconstructions of Schove (1979, 1983), demonstrates that this is exactly the case during the Maunder minimum: the amplitude of the cycle is positively correlated to the rise time with r=0,84 and p=0.037.

Therefore, the relation between the cycle rise time and amplitude is a consequence of the dependence of the cycle amplitude on the speed of the deep meridional circulation. The Waldmeier rule (faster rise = stronger cycle) is a manifestation of this relation in the case of diffusion-dominated regime, and it does not hold in the advection-dominated regime.

### 4.6 Sunspot cycle memory

Charbonneau and Dikpati (2000) suggested that the duration of the Sun's memory of its own magnetic field in flux-transport dynamos is governed primarily by the meridional flow speed, and is no less than two solar cycles. This "memory" manifests in the correlations between the parameters of cycle n and the amplitude of cycles n+1, n+2, etc. The correlation between Vdeep and the amplitude of the solar cycle for the last 140 years shown in Fig.7a demonstrates that the sunspot cycle has a short memory. The sunspot cycle maximum is correlated to the speed of the deep circulation preceding it, with a smaller contribution from the previous cycle, and no contribution from earlier cycles.

The one-cycle only memory is in agreement with the predictions of the model of Yeates et al. (2007) that in the diffusion-dominated regime the toroidal field for cycle n+1 is produced primarily from the poloidal field of cycle n, with a small contribution from cycle n-1, while in the advection-dominated regime, poloidal fields from cycles n, n-1, and n-2 combine to produce the toroidal field for cycle n+1. This can be understood as a consequence of the unidirectional correlation between Vsurf and Vdeep: the poloidal field $Bpol_n$ is formed on the decline phase of cycle n from the toroidal field of cycle n, $Btor_n$ which is in turn correlated to the speed of the deep circulation preceding it, $Vdeep_n$, and on the speed of the surface circulation $Vsurf_n$; $Bpol_n$ is the seed for the toroidal field of cycle n+1, $Btor_{n+1}$ which obviously depends on $Bpol_n$ and thus indirectly on $Vdeep_n$ through the dependence of $Bpol_n$ on $Btor_n$, and depends also on the the speed of the deep circulation $Vdeep_{n+1}$; $Vsurf_n$ and $Vdeep_{n+1}$ are highly correlated but $Vdeep_n$ and $Vsurf_n$ are uncorrelated, so the correlation between $Btor_{n+1}$ and $Vdeep_n$ only exists because it is mediated by the correlation chain Vdeep$_n$ → Btor$_n$; Btor$_n$ + Vsurf$_n$ → Bpol$_n$; Bpol$_n$ + Vdeep$_{n+1}$ → Btor$_{n+1}$, but it cannot survive longer because the chain breaks between Vdeep and Vsurf.

However, the second half of the prediction – about the long memory in the advection-dominated regime – is not confirmed by our results. The cross-correlation between the maximum sunspot number and the speed of the deep meridional circulation for the period of the Maunder minimum (Fig.8c) demonstrates that the memory in the advection-dominated regime is equally short as in the diffusion-dominated regime.

The reason for the correlations to persist from cycle to cycle (and therefore to provide longer-than-one-cycle memory) in the advection-dominated regime and not in the diffusion-dominated regime in the model of Yeates et al. (2007) is that in the former case fluctuations are passed on in both the poloidal-to-toroidal and in the toroidal-to-poloidal phases of the cycle by means of the meridional circulation, while in the latter case the poloidal field is carried to the tachocline by means of downward diffusion "short-circuiting" the meridional circulation, so the correlations are passed on only in the poloidal-to-toroidal phase. But the time between the poloidal field maximum and the next toroidal field maximum is comparable in the two regimes, and the estimated speed of the deep meridional circulation is in agreement with the speed derived from the sunspot butterfly diagram (Hathaway et al., 2003). Therefore it doesn't seem probable that in the diffusion-dominated regime the downward diffusion "short-circuits" the meridional circulation. Rather, we could suggest that in both regimes the flux makes one full turnover per cycle carried by the meridional circulation, with the speed of the deep circulation determining the cycle period, and the diffusivity determining the relation between the speed of the deep circulation and the cycle amplitude.

Alternatively, in the advection dominated model of Charbonneau and Dikpati (2000), the longer memory is achieved through slow deep circulation combined with survival of multiple old-cycle polar fields thanks to the low diffusivity during the prolonged time for advective transport from surface high latitudes to deep sunspot latitudes. But it should be noted that the extremely low value of the deep circulation speed assumed by Charbonneau and Dikpati (2000) and apparently in all subsequent modifications of this model - 0.15 m/s versus 15 m/s for the surface meridional circulation – is quite speculative, and is an order of magnitude smaller than the speed estimated from the butterfly diagram (Hathaway et al., 2003). The authors themselves stress that the two-cycle lag which they find is a property of the model, and that adopting a different meridional flow pattern might well yield shorter or longer lags. As commented in 3.3, the ratio of the speeds of the deep and surface circulations determines the depth where the circulation reverses from poleward at the surface to equatorward below. A surface circulation two orders of magnitude faster than the deep circulation would mean a reversal depth at 0.95 $R_S$. Giles et al. (1998) from time-distance helioseismology did evaluate a reversal depth of 0.94 $R_s$ but Braun and Fan (1998) pointed out that the above result is not consistent with the observations of SOHO Solar Investigations Observation – Michelson Doppler Imager (SOI-MDI). Later Giles (2000), again from time-distance helioseismology, found a reversal depth of 0.8 $R_s$, while Duvall and Kosovichev (2001) found no reversal at all down to 0.725 $R_s$. Krieger et al. (2007), applying the Fourier-Hankel spectral decomposition method of Braun and Fan (1998) on data from SOHO SOI-MDI for 1999, also found a reversal depth of 0.8 $R_s$. In contrast, recently Mitra-Kraev and Thompson (2007), again from analysis of SOHO SOI-MDI data, reported flow reversal at about 0.95 $R_S$ with a possible deeper counter-cell (however, Jouve and Brun, 2007, demonstrated that it is unlikely that multicellular meridional flows persist for a long period of time inside the Sun). As seen, the estimations of the reversal depth are quite controversial and to the best of our knowledge, the initial estimations of Giles et al. (1998) and the observation of Mitra-Kraev and Thompson (2007) are so far the only results indicating such a shallow reversal depth which would allow deep circulation as slow as assumed in the model of Charbonneau and Dikpati (2000). But apart from the fact that such a slow deep circulation contradicts the estimations from the butterfly diagram (Hathaway et al., 2003), such a shallow reversal depth would mean high average diffusivity in the part of the convective zone involved in equatorward motion, and high diffusivity combined with low circulation speed would mean diffusion-dominated regime where the old-cycle polar fields could not survive long enough to introduce a long memory. We therefore feel that neither such a low speed of the deep equatorward circulation as assumed by Chabonneau and Dikpati (2000), nor such a shallow reversal depth as such slow deep circulation would mean, are very realistic.

5 Prediction for solar cycle 24

Recently Kane (2007a) made an extensive review of the methods for predicting future sunspot cycles, and summarized more than 40 predictions for solar cycle 24. He divided them into four broad groups: predictions based on (1) physical analysis including precursor methods, predicting maximum sunspot number for the next cycle between 45 and 180; (2) spectral analysis predicting values from 43±34 to 180; (3) statistical analysis including neural networks: predictions from below 50 to 190, the latter if the cycle is a fast riser; (4) other predictions available as abstracts or private communications: from 70±18 to 163±37. We should note here that some of the predictions included in group (4) are by now already published: Choudhuri et al. (2007), Hiremath (2007a), Tlatov (2007). To these we can add a few more predictions published after Kane's (2007a) paper: Jiang et al. (2007), using Babcock-Leighton mechanism with high diffusivity of order $10^{12}$ cm$^2$s$^{-1}$, predict cycle 24 to be the weakest cycle in a long time; Quassim et al. (2007), based on spectral analysis of the three-century long sunspot number data-set, predict cycle 24 to reach a maximum sunspot number of 110 in 2011; Kane (2007b), based on the correlation between the number of sunspot groups during a sunspot minimum year in the latitude belt 0-30°, predicts the maximum sunspot number in cycle 24 to be 129.7±16.3; Kane (2007c), from spectral analysis by the maximum entropy method, found extrapolated Rz(max) in the ranges 80 – 101 (mean 92) for cycle 24 during years 2011 – 2014; Callebaut and Makarova (2007), from the polar faculae in the two hemispheres predicted the times of peaks in Wolf numbers. Hiremath (2007b) modeling the sunspot cycle as a damped and forced harmonic oscillator, predicted 110 and 9.34 as the amplitude and period of cycle 24, respectively.

In summary, the predictions vary by more than a factor of 4: from 45 to 190. They are based on combinations of input parameters from virtually the same data set, but on different hypotheses about the operation of the solar dynamo. Therefore, such predictions, apart from their potential practical significance, are an important test for the underlying hypotheses.

It is not the purpose of this paper to give one more prediction. But the relations found from the estimated long-term variations in the meridional circulation do have some predictive power, so we will demonstrate what follows from these relations.

As shown in the previous section, we find that recently solar dynamo has been operating in diffusion-dominated regime, and that fluctuations are passed on in the poloidal-to-toroidal phase of the cycle. In this regime, we find that the time from sunspot maximum to the highest aa-index maximum on the sunspot decline phase ("delay-1") is correlated with the amplitude and time of the maximum of the following sunspot cycle. The dependencies are pretty linear (Fig.5b and Fig.6), and are expressed as:

(sunspot max)$_{n+1}$ = 70.9+1.4727*(delay-1)$_n$  (1)

(delay+1)$_{n+1}$ = 139-1.268*(delay-1)$_n$   (2)

(1) gives 125±17 as the expected maximum sunspot number for cycle 24, and (2) gives 92±8 months between the last aa-index maximum (August 2003) and the next sunspot maximum, which means sunspot maximum in April 2011 ± 8 months.

## 6 Summary and conclusion

We have used the data series of sunspot numbers and group sunspot numbers, aa-index of geomagnetic activity and galactic cosmic rays minima from cosmogenic isotopes to estimate the long-term variations in the solar meridional circulation, and to investigate the relation between meridional circulation and solar activity. We assume a flux-transport dynamo in which the meridional circulation is a key factor determining the cycle amplitude and period, and we assume that this meridional circulation completes one full turnover per cycle. Under these assumptions we derive the speed of the surface and of the deep meridional circulation and the depth where the circulation reverses from poleward at the surface to equatorward below for the last 13 sunspot cycles and in the period of the Maunder minimum. Our results can be summarized as follows:

1. The surface poleward circulation, the deep equatorward circulation, and the reversal depth, all have long-term variations related to the long-term variations in solar activity.

2. Since 1868, the amplitude of the sunspot cycle has been positively correlated to the speed of the deep circulation preceding it which, according to the model of Yeates et al. (2007), means that in the last 140 years the solar dynamo has been operating in diffusion-dominated regime.

3. During the Maunder minimum in the 17$^{th}$ century, the amplitude of the sunspot cycle was negatively correlated to the speed of the deep circulation preceding it, therefore during the Maunder minimum the solar dynamo operated in advection-dominated regime.

4. The speed of the deep circulation determines the rise time of the sunspot cycle. This, together with the correlation between the speed of the deep circulation and the cycle amplitude predicted by the model of Yeates et al. (2007) and confirmed by our results, explains the relation between the cycle's rise time and amplitude The Waldmeier rule (faster rise = higher cycle maximum) is a manifestation of this relation in the diffusion-dominated regime, while in the advection-dominated regime the relation is opposite (faster rise = lower cycle maximum).

As shown by Cameron and Schüssler (2007) and Schüssler (2007), the relation between the speed of the deep circulation and the cycle amplitude, together with the overlapping of cycles also explain other well known statistical characteristics of the solar cycle since the end of the Maunder minimum: that large-amplitude cycles are preceded by high minima, and that large-amplitude cycles are preceded by short-period cycles (e.g. Hathaway and Wilson, 2004).

5. In both regimes, the "memory" of the sunspot cycle is short, one-only cycle, implying that diffusion plays an important role even in the advection-dominated regime.

6. The speed of the surface meridional circulation is correlated to the amplitude and rise time of the following sunspot cycle, and can be used as a precursor for the following cycle.

7. Our results confirm the validity of the Babcock-Leighton dynamo mechanism, and the model of Yeates et al. (2007).

## Future work

Our primary goal in this study was to estimate the long-term (of the order of the secular solar cycle) variations in the meridional circulation. To highlight these long-term variations, in Figs. 1, 2, 3, 4, and 8, the 3-point moving averages are given which smooth the differences between odd and even cycles. Actually, there is a systematic difference in the circulation speeds between odd and even cycles corresponding to the difference between the cycles' amplitudes: both the deep circulation preceding the cycle maximum, and the surface speeds following it, are as a rule faster in odd than in even cycles. The cases of violation of this rule correspond to the cases of violation of the Gnevyshev-Ohl rule. Further, here we have studied the meridional circulation and the solar activity averaged for the two solar hemispheres, while both circulations have different evolution in the southern and in the northern hemispheres (Dikpati et al., 2007). These two points are being investigated now and will be the subject of a following paper.

As mentioned in the Introduction, the long-term variations in the correlation between sunspot and geomagnetic activity can be due to either the changing lag between the sunspot maximum and the peak in geomagnetic activity on the sunspot cycle decline phase, or to the changing of the relative importance of the two peaks of geomagnetic activity in the sunspot cycle, or to both. Here we have investigated the first cause. To determine the long-term variations in the relative importance of the geomagnetic activity peaks related to the two types of solar activity, the method proposed by Feynman (1982) can be used to decompose the solar magnetic field into a poloidal and a toroidal components. We have done this and have found that these two components have different long-term evolution. Our preliminary results indicate that the long-term variations in terrestrial surface air temperature are highly correlated to the long-term variations in the solar poloidal field while the sunspot number is proportional to the solar toroidal field, so using the sunspot number to evaluate solar influences on climate leads to the underestimation of the role of Sun for global climate change. This question will be a subject of a separate paper.

## Acknowledgements

This study was partly supported by EOARD grant 063048.